# Dynamical mean-field theory study of a ferromagnetic $CrI_3$ monolayer

Chang-Jong Kang,[1,2,†] Jeonghoon Hong,[3] and Jeongwoo Kim[3,4,*]

[1]*Department of Physics, Chungnam National University, Daejeon 34134, Korea*

[2]*Institute of Quantum Systems, Chungnam National University, Daejeon 34134, Korea*

[3]*Department of Physics, Incheon National University, Incheon 22012, Korea*

[4]*Intelligent Sensor Convergence Research Center, Incheon National University, Incheon 22012, Korea*

corresponding author e-mail

†:cjkang87@cnu.ac.kr; *:kjwlou@inu.ac.kr

**Abstract**

We have employed one of the well-known many-body techniques, density functional theory plus dynamical mean-field theory (DFT + DMFT), to investigate the electronic structure of ferromagnetic monolayer $CrI_3$ as a function of temperature and hole-doping concentration. The computed magnetic susceptibility follows the Curie's law, indicating that the ferromagnetism of monolayer $CrI_3$ originates from localized magnetic moments of Cr atoms rather than Stoner-type itinerant ones. The DFT + DMFT calculations show a different coherent temperature for each spin component, demonstrating apparent strong spin-dependent electronic correlation effects in monolayer $CrI_3$. Furthermore, we have explored the doping-dependent electronic structure of monolayer $CrI_3$ and found that its electronic and magnetic properties is easily tunable by the hole-doping.

## 1. Introduction

Atomically thin CrI$_3$ is one of the representative two-dimensional (2D) magnets with unique electronic/magnetic properties and their potential applications [1–4]. Such a long-range magnetic order is stabilized by a sizable magnetic anisotropy driven by spin-orbit coupling [5] because it is not allowed in the 2D Heisenberg model according to the Mermin-Wagner theorem [6]. Among several van der Waals ferromagnets with atomic thickness, CrI$_3$ is special because of the coexistence of its insulating nature and magnetic tunability [7–9]. For example, the inter-layer magnetic order is determined by the stacking sequence of a few-layer CrI$_3$ [10,11], leading to the reversible magnetic transition between ferromagnetic and antiferromagnetic states [12]. In addition, its magnetic anisotropy, which plays a central role in controllability of magnetic devices, can be changed by various routes such as doping, heterostructure with graphene, and external irradiation [13].

Nevertheless, until now, correlation effect on 2D magnets has not been thoroughly investigated. The restricted geometry of low dimensions induces quantum confinement of electrons. Furthermore, the reduced dimensionality tends to enhance electron-electron Coulomb interactions due to the diminished electronic screening effect [14]. Therefore, it is required to explore whether the interesting 2D magnetism is affected by correlation effect.

In this article, we study temperature- and doping-dependent electronic/magnetic properties of monolayer CrI$_3$ using the density functional theory plus dynamical mean-field theory (DFT + DMFT) approach. The calculated magnetic susceptibility obeys the Curie's law and yield an effective local magnetic moment of ~3.4 $\mu_B$, indicating that the ferromagnetism in monolayer CrI$_3$ originates from the localized magnetic moments rather than Stoner-type itinerant ones. Temperature-dependent DMFT calculations exhibits different coherent temperatures $T^*$ for spin-up/down components, implying strong spin dependent electronic correlation effects in monolayer CrI$_3$. We also investigate the doping-dependent electronic structure of monolayer CrI3 and discuss the tunability of its electronic and magnetic properties by the hole-doping.

## 2. Methods

Fully charge self-consistent DFT+DMFT calculations [15–17] implemented in Wien2k package [18] were performed with formalisms described in Ref. [19]. The relaxed crystal structure of monolayer CrI3 was adopted for our study. In DFT calculations, the generalized gradient approximation (GGA) of Perdew−Burke−Ernzerhof (PBE) was used for the exchange-correlation functional. The muffin tin radii were chosen to be 2.50 Bohr radii for both Cr and I; the size of a plane-wave basis set was determined from $R_{mt}K_{max}$ of 7.0, where $R_{mt}$ is the smallest atomic muffin tin radius and $K_{max}$ is the largest planewave vector. A 23 × 23 × 9 k-point mesh was used for the Brillouin zone integration. In DMFT calculations, the continuous-time quantum Monte Carlo method (CTQMC) [20,21] was implemented for a local impurity solver. We choose a wide hybridization energy window from −10 to 10 eV with respect to the Fermi level $E_F$. The Ising type interaction form was applied for a local Coulomb interaction Hamiltonian with on-site Coulomb repulsion ($U$ = 10 eV) and Hund's coupling ($J_H$ = 1 eV), where the chosen $U$ and $J_H$ values were adequate for insulating 3$d$ transition-metal compounds. The exact double counting (DC) scheme devised by Haule was adopted, which eliminates the DC issues in correlated materials [22]. The maximum entropy method [23] was used for analytical continuation to obtain the electronic self-energy due to electron-electron interaction on the real frequency. In order to simulate hole-doping effects in the DMFT calculations, a positive background charge was introduced. It is namely the virtual crystal approximation and has been successfully applied to other correlated electron systems within the DMFT framework [24,25].

## 3. Results

We investigate the effect of dynamical correlations on the magnetic and electronic properties of monolayer CrI$_3$. Using the DFT+DMFT method, we calculate the magnetic moment of monolayer CrI$_3$ varying temperature from 116 to 232 K [Fig. 1(a)]. In this study, we focus on qualitative trends of magnetic/electronic characteristics upon temperature change rather than a quantitative comparison between experiment and theory, because the Curie temperature is overestimated in the Ising-type interaction [26–28] and non-local correlations are neglected in the DMFT scheme. Note that the similar DFT + DMFT method describes the electronic structure of bulk CrI$_3$ well compared to experiments [29]. Hence, the DFT + DMFT method is enough to explore the electronic structure of monolayer CrI$_3$. Non-zero magnetization emerges at ~190 K (computed Curie temperature), as expected much greater than the experimental Curie temperature (~45 K) [1], and becomes an ideal value of 3.0 $\mu_B$ at a low temperature regime (T = 116 K). The effective local magnetic moment of ~3.4 $\mu_B$ is estimated from the computed magnetic susceptibility [Fig. 1(b)] that follows the Curie's law. It indicates that the primary source of the ferromagnetism in monolayer CrI$_3$ is the localized magnetic moments rather than Stoner-type itinerant one. The total occupation number of Cr 3$d$ orbital is 4.12 in our DMFT calculations, which is different from the occupation number of 3 estimated from the ionic picture of Cr$^{3+}$ systems. Due to the strong hybridization between Cr-$e_g$ and I-$p$ orbitals, an additional electron occupation is induced in Cr $e_g$, leading to the deviation from the ionic value. The total occupation number is temperature independent, and comparable to our DFT results. Note that the ordered magnetic moment at the low temperature (T = 116 K) estimated from the DMFT calculations is almost same as our DFT results, implying that correlation effects do not affect the size of the ordered magnetic moment in two-dimensional magnet CrI$_3$.

Magnetism of monolayer CrI$_3$ affects its electronic structure as shown in Fig. 1(c) and

Fig. 2, in which ferromagnetism and paramagnetism are realized at low and high temperatures, respectively, and the electronic structure is temperature dependent. Occupied states near the Fermi level mostly originate from I atoms and, therefore, our analysis is concentrated on unoccupied states near the Fermi level. Above the Curie temperature (T = 232 K), both spin-up and down components have identical density of states [Fig. 1(c)], indicating no net magnetization. For both spin components, Cr-$e_g$ bands above the Fermi level are identical and incoherent (Fig. 2). Below the Curie temperature (T < 190 K), the Cr-$e_g$ bands exhibit different behaviors depending on their spin component: the spin-up density of states (DOS) for Cr-$e_g$ becomes localized and eventually show a large coherent peak with decreasing temperature, while the spin-down DOS for Cr-$e_g$ are extended over a broad energy region even at the low temperature [Fig. 1(c)]. The disparate trends of the spin-up/-down Cr-$e_g$ states can also be found in the band dispersion: The spin-up bands of Cr-$e_g$ become coherent quasi-particle at T= 116 K, while the spin-down bands fade away upon cooling. (Fig. 2). Similarly, the spin-up component of Cr-$t_{2g}$ states (~-1 eV) forms occupied quasi-particle bands with decreasing temperature (Fig.2) whereas unoccupied spin-down Cr-$t_{2g}$ states (~1.7 eV) are less coherent even at the low temperature [Fig.1 (c)]. Consequently, these differences between two spin components explain the spin imbalance and the net magnetization presented in Fig. 1(a). Note that the coherent temperatures $T^*$ for the spin-up and down components are different and $T^*$ for the spin-up is higher than that for the spin-down. It distinctly indicates the strong spin dependence of correlation effects in monolayer $CrI_3$.

To elucidate how spin and orbital degrees of freedom are intertwined, we compute the spin-resolved self-energy of Cr-$t_{2g}$ and $e_g$ orbitals in monolayer $CrI_3$. Figure 3 shows the imaginary part of the self-energy on the Matsubara frequency axis and the quasi-particle weight Z as a function of temperature. Z is evaluated as $1/Z = 1 - \partial \text{Im}\Sigma(i\omega)/\partial\omega|_{\omega \to 0^+}$, comparable to the result from analytical continuation. As displayed in the imaginary part of the self-energy

[Fig. 3(a-d)], the spin dependency does not occur above the Curie temperature (≥190 K) for Cr-$t_{2g}$ and $e_g$ orbitals. As temperature decreases below the Curie temperature, the self-energies start to differ from each spin component. Such trend is also shown in the quasi-particle weight $Z$ [Fig. 3(e)]. $Z$ values for spin-up (solid lines)/-down (dashed lines) states are identical at high temperatures, but they are separated below the Curie temperature.

Interestingly, Cr-$e_g$ spin-down states exhibit an unusual trend that $Z$ decreases with lowering temperature [Fig. 3(e)]. $Z$ value normally increases at low temperature, demonstrating that correlated orbitals become quasi-particle bands. The Cr-$t_{2g}$ and $e_g$ spin-up states show the aforementioned trend, indicating the presence of the Landau-Fermi liquid phase at low temperature. On the contrary, Z of the $e_g$ spin-down declines with decreasing temperature. It indicates that the Cr-$e_g$ spin-down states experience a larger Coulomb interaction than its spin-up counterpart when the system has net magnetization. As a result, the coherent temperatures $T^*$ for Cr-$e_g$ spin-up and down states are different and $T^*$ for Cr-$e_g$ spin-down is smaller than that for Cr-$e_g$ spin-up states. The differentiation is mainly due to the spin exchange interaction.

We also investigate the doping-dependent electronic structure of monolayer CrI$_3$ within the DMFT method. We compute the Cr-3$d$ occupancy varying the hole-doping concentration at T = 116 K below the Curie temperature. As the doping concentration increases from 0 to 0.3 e/cell, the occupancy is reduced from 4.14 to 4.116 [Fig. 4(a)]. Because the iodine valence band states near the Fermi level are mainly affected by the hole doping [Fig. 4(c)], a change in the Cr-3$d$ occupancy is relatively small. On the contrary, the magnetization of Cr 3$d$ is very sensitive to the hole-doping concentration: an increase of the magnetization roughly corresponds to the hole-doping concentration. Since the number of occupied Cr-3$d$ states is almost kept constant, the noticeable increase of magnetization can be attributed to the change in the electron configuration: the occupancy of spin-up Cr-$e_g$ states. By the hole doping, a pronounced peak of spin-up Cr-$e_g$ states emerges around -2.2 eV [Fig. 4(c)], which is the chief reason for the variation of the magnetization. Above the Fermi level (~2.5 eV), another new spin-down $e_g$ peak appears, which is absent in the pristine case. Therefore, we can conclude that the hole-doping is an effective way to manipulate the magnetic and electronic properties of monolayer $CrI_3$.

## 4. Conclusion

We investigated temperature and doping dependent electronic and magnetic properties of monolayer $CrI_3$ within the DFT + DMFT method. The computed magnetic susceptibility follows the Curie's law with the effective local magnetic moment of ~3.4 $\mu_B$, indicating that the ferromagnetism in monolayer $CrI_3$ originates from localized magnetic moments rather than Stoner-type itinerant ones. The computed Curie temperature is ~190 K, which is much greater than the experimental value due to the Ising-type Coulomb interaction adopted in the DMFT calculations. Temperature-dependent DMFT calculations show different coherent temperatures $T^*$ for two spin components and $T^*$ for the spin-up is higher than that for the spin-down component, demonstrating the strong spin dependent electronic correlation effects in monolayer $CrI_3$. We also explored doping dependent electronic structure of monolayer $CrI_3$ and found that its electronic and magnetic properties is easily tunable by the hole doping.

**Acknowledgement**

We are grateful to G. L. Pascut for useful discussions. This work was supported by Incheon National University Research Grant in 2021 (2021-0013).

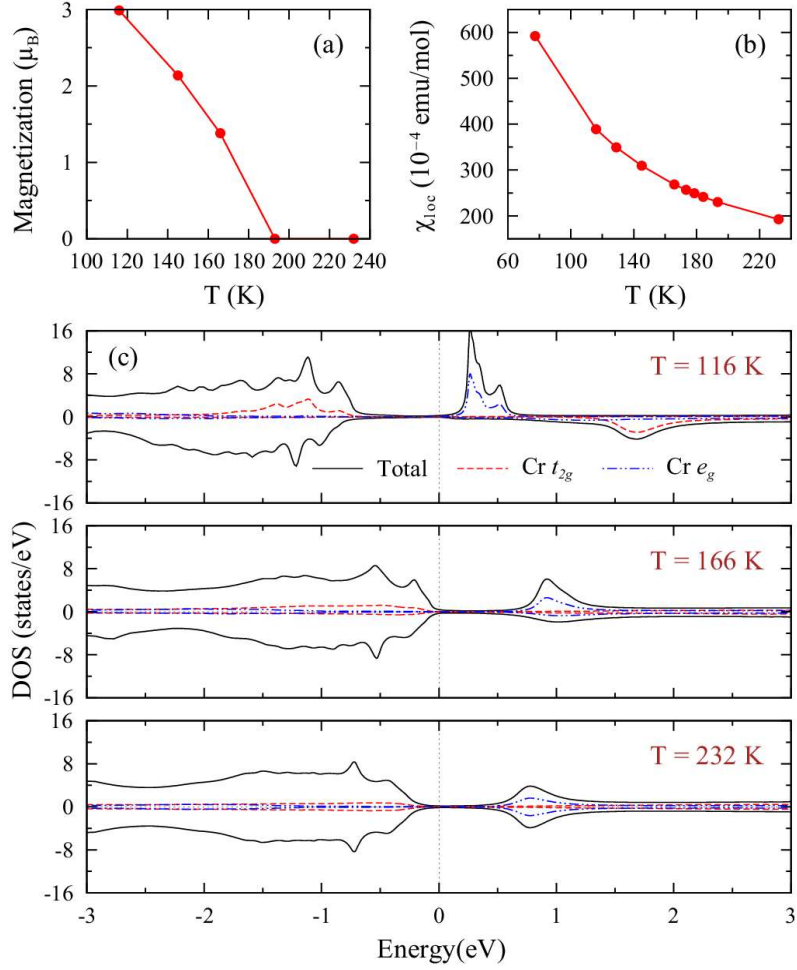

Figure 1. DMFT calculations in monolayer CrI$_3$. (a) Magnetization as a function of temperature. The Curie temperature is ~190 K. (b) Local magnetic susceptibility $\chi_{loc}$ as a function of temperature. The red lines in (a) and (b) are guide for eyes. (c) Temperature dependent density of states calculated from DMFT calculations. The chemical potential is set to be zero in the density of states. Red and blue lines correspond to Cr d-t2g and d-eg orbital characters, respectively.

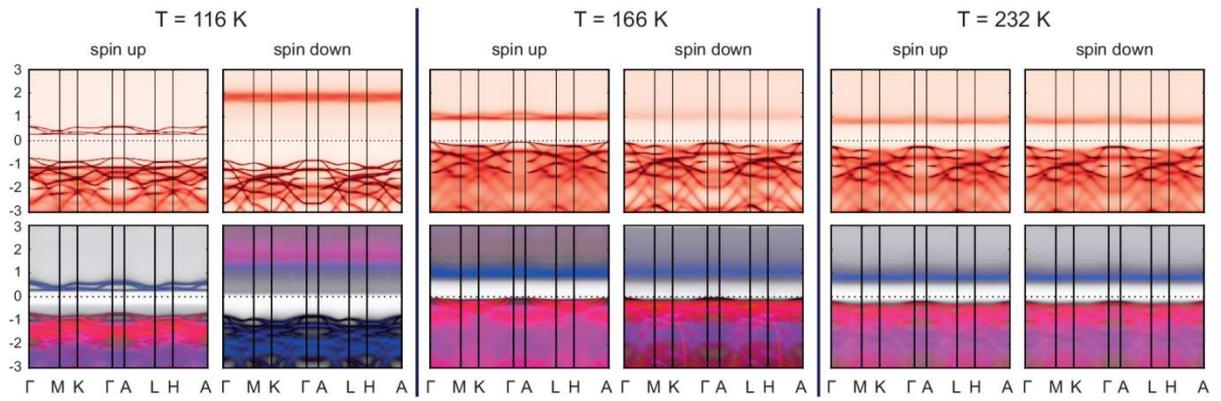

Figure 2. Band dispersion of monolayer CrI$_3$ computed within DFT + DMFT. Momentum($k$)-resolved spectral functions A($k$, $\omega$) for each spin up and down component at T = 116 K (Left), T = 166 K (Middle), and T = 232 K (Right). Cr d-t2g and Cr d-eg orbital characters are presented by red and blue, respectively, in the bottom figure for each spin component and temperature.

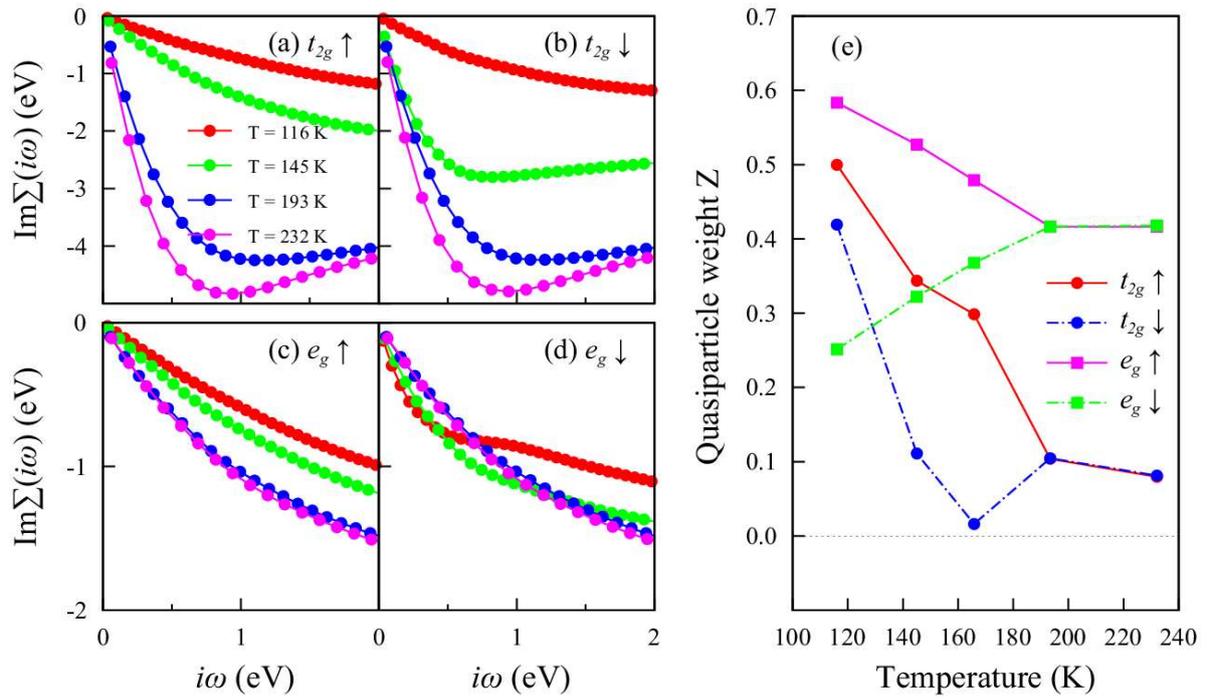

Figure 3. DMFT self-energy. (a-d) Imaginary part of self-energy of Cr-t2g and Cr-eg orbitals for both spin configurations. (e) Temperature dependent quasi-particle weight Z obtained from DMFT calculations. The Cr-$e_g$ orbital has higher Z values than the $t_{2g}$ orbital over the temperature range, indicating that correlation effect in the $t_{2g}$ orbital is more dominant than in $e_g$ orbital.

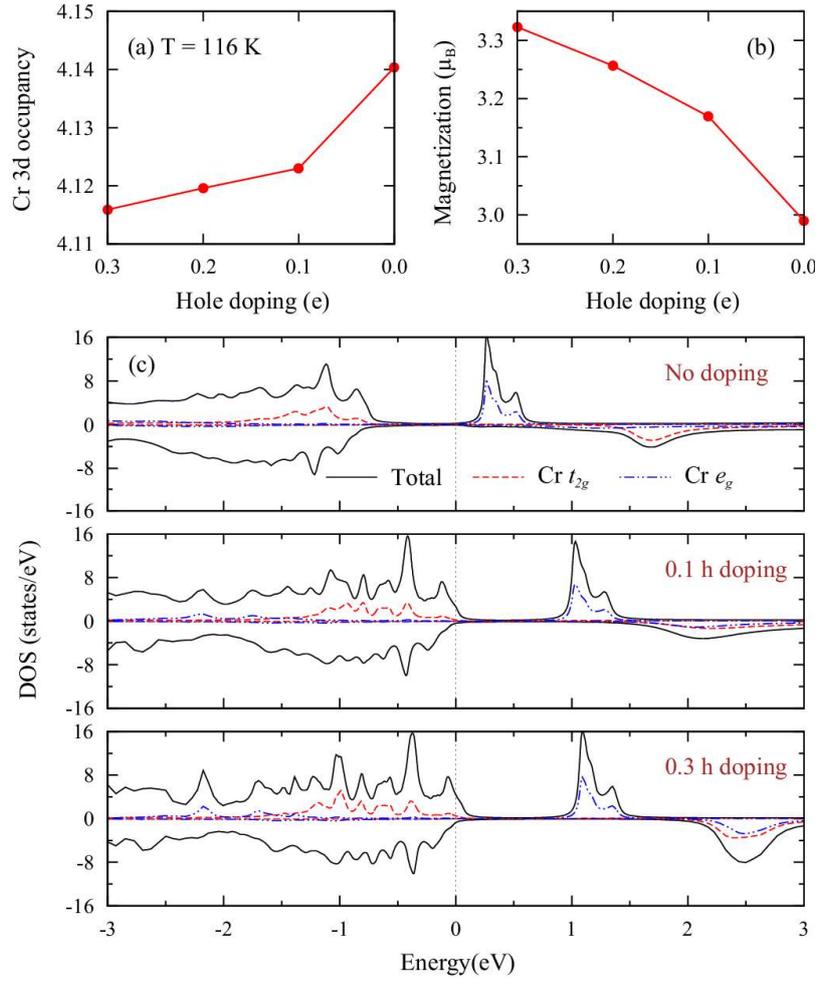

Figure 4. Doping-dependent DMFT calculations of monolayer CrI$_3$ at T = 116 K. (a) Cr 3d occupancy as a function of hole-doping concentration. (b) Magnetization as a function of hole-doping concentration. The red lines in (a) and (b) are guide for eyes. (c) Doping dependent density of states calculated from DMFT calculations. The chemical potential is set to be zero in the density of states. Red and blue lines correspond to Cr d-t2g and d-eg orbital characters, respectively.